\documentclass[conference]{IEEEtran}
\usepackage{cite}
\usepackage{graphicx} 
\usepackage[left=0.625in,right=0.625in,top=0.75in,bottom=1in]{geometry}
\usepackage{subcaption}
\usepackage{algorithm}
\usepackage{algpseudocode}
\usepackage{indentfirst}
\usepackage{amsmath}
\usepackage{amsfonts}
\usepackage{amssymb}
\usepackage{xcolor}
\def\BibTeX{{\rm B\kern-.05em{\sc i\kern-.025em b}\kern-.08em
    T\kern-.1667em\lower.7ex\hbox{E}\kern-.125emX}}
\usepackage{subcaption}
\usepackage{multirow}
\usepackage{booktabs}
\usepackage{adjustbox}
\usepackage{pifont}
\usepackage{float}
\usepackage{url}
\usepackage{enumitem}
\usepackage[font=small]{caption}
\usepackage[utf8]{inputenc}

\begin{document}
\title{Leveraging MoE-based Large Language Model for Zero-Shot Multi-Task Semantic Communication}
\author{
    \IEEEauthorblockN{Sin-Yu Huang, Renjie Liao, and Vincent W.S. Wong}
        Department of Electrical and Computer Engineering, The University of British Columbia, Vancouver, Canada\\
        Email: \{syhuang, rjliao, vincentw\}@ece.ubc.ca
}
\maketitle
\begin{abstract}
    Multi-task semantic communication (SC) can reduce the computational resources in wireless systems since retraining is not required when switching between tasks. However, existing approaches typically rely on task-specific embeddings to identify the intended task, necessitating retraining the entire model when given a new task. Consequently, this drives the need for a multi-task SC system that can handle new tasks without additional training, known as zero-shot learning. Inspired by the superior zero-shot capabilities of large language models (LLMs), we leverage pre-trained instruction-tuned LLMs, referred to as fine-tuned language net (FLAN), to improve the generalization capability. We incorporate a mixture-of-experts (MoE) architecture in the FLAN model and propose MoE-FLAN-SC architecture for multi-task SC systems. Our proposed MoE-FLAN-SC architecture can further improve the performance of FLAN-T5 model without increasing the computational cost. Moreover, we design a multi-task feature extraction module (FEM) which can adaptively extract relevant features across various tasks given the provided features and signal-to-noise ratio (SNR). Simulation results show that our proposed MoE-FLAN-SC architecture outperforms three state-of-the-art models in terms of the average accuracy on four different unseen tasks.
\end{abstract}

\section{Introduction}
\label{Sec:intro}

Task-oriented semantic communication (SC) is an emerging technology for the sixth-generation (6G) wireless networks~\cite{Deniz2023}. In task-oriented SC, the transmitter only sends the task-related features to the receiver. In~\cite{Shao2022}, Shao \textit{et al.} proposed a learning-based task-oriented communication scheme that can efficiently extract features from images for image classification tasks. In~\cite{Hu2023}, Hu \textit{et al.} studied the impact of semantic noise on task-oriented SC systems and introduced a feature importance module that can dynamically adjust the weights for different features. Nevertheless, the aforementioned works only study a single-task scenario. To perform a new task, the entire model needs to be retrained.

In~\cite{Zhang2024}, U-DeepSC is proposed to handle multiple tasks across three modalities: text, speech, and image. It can achieve performance comparable to that of single-task SC systems trained individually on each task. However, for each task, U-DeepSC relies on a task-specific embedding vector shared by the transmitter and receiver in order to identify the tasks to be performed. However, it limits the model’s generalization ability to tasks that have not been explicitly trained on, referred to as unseen tasks~\cite{Wang2024}. On the other hand, some recent works have proposed approaches to transfer knowledge from an existing task to an unseen task. In~\cite{Leung2024}, Leung \textit{et al.} proposed a generalizable multi-task communication paradigm, where only the decoder needs to be retrained for unseen tasks. However, retraining the decoder, which is nearly half of the model size, is both time- and resource-intensive.

Performing unseen tasks without either additional training or specific examples for those tasks is called zero-shot learning. In this context, large language models (LLMs) have shown to possess strong zero-shot ability. In~\cite{Raffel2020}, Raffel \textit{et al.} introduced a unified text-to-text transfer Transformer (T5) model, showcasing LLMs' strong transfer learning capabilities across tasks such as summarization, question answering, and text classification. Building on the T5 model, Chung \textit{et al.} in~\cite{Chung2024} proposed the fine-tuned language net (FLAN)-T5 model, demonstrating that instruction fine-tuning with a number of tasks can significantly enhance the model’s zero-shot performance. Despite the impressive results achieved by LLMs, their performance is at the expense of larger model sizes.

In multi-task SC systems, another challenge arises due to gradient conflicts between tasks~\cite{Yu2020}. During the training process, the gradients of different tasks can point to different directions. Therefore, updating the model parameters based on gradients of one specific task can degrade the performance of other tasks. Recall that task-oriented SC transmits only the task-relevant features to complete the intended task. In~\cite{Zhang2024}, U-DeepSC uses a separate feature selection module (FSM) to select task-oriented features for each task. This approach effectively avoids gradient conflicts since each module independently optimizes the feature selection process specific to its task. However, when the number of tasks increases, the number of FSMs increases linearly, resulting in a larger model size. Furthermore, when encountering an unseen task, a new FSM needs to be trained. 
For a multi-task SC system, we face the following questions:
\begin{enumerate}[label=Q\arabic*:]
    \item \textit{How to design a multi-task SC system that can effectively tackle those unseen tasks without retraining?}
    \item \textit{How to satisfy the diverse task requirements and improve the performance in a multi-task SC system without increasing the computational cost?}
    
\end{enumerate}
\begin{figure*}[t]
\centering
\includegraphics[width=0.95\textwidth]{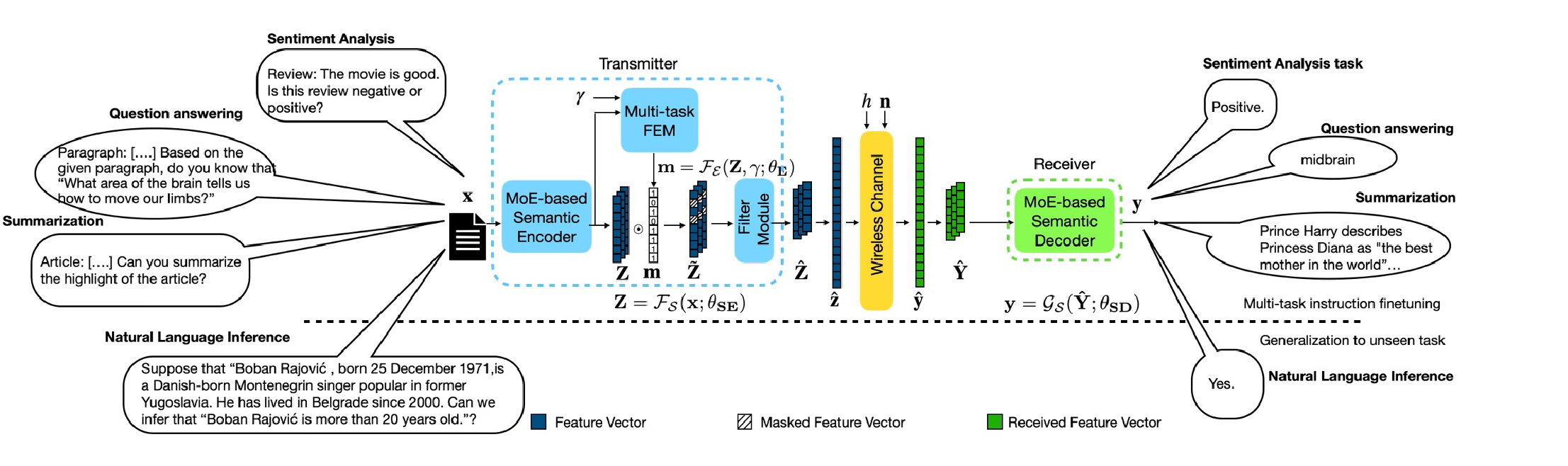}
\caption{\small The system model of the proposed MoE-FLAN-SC architecture, which is trained on multiple natural language processing tasks, including sentiment analysis, question answering, and summarization, using natural language prompts. By training on a diverse range of tasks, the model enhances its ability to generalize to previously unseen tasks.}
\label{Fig:system_model}
\end{figure*}

To address the above questions, in this paper, we leverage the mixture-of-experts (MoE)-based LLM~\cite{Shen2023} in a multi-task SC system. 
The model aims to perform zero-shot learning for unseen tasks while filters out the task-irrelevant features for each task.
The contributions of this paper are as follows:
\begin{itemize}
    \item We propose an MoE-FLAN-SC architecture that facilitates zero-shot learning for multi-task SC system. The proposed architecture integrates MoE in a pretrained FLAN-T5 model and uses a single multi-task feature extraction module (FEM) to adaptively filter out irrelevant features for all tasks, thereby reducing the transmission overhead. We propose a two-phase algorithm to instruction-tune the model using public pool of prompts (P3)~\cite{Sanh2022}, which is a collection of datasets with prompt templates.
    \item The proposed MoE-FLAN-SC architecture can perform zero-shot learning on various unseen tasks. By integrating MoE into our model, it improves zero-shot performance with the same floating point operations per second (FLOPs) per token. The multi-task FEM also uses the concept of MoE that includes multiple extractor modules to filter out irrelevant features under different signal-to-noise ratio (SNR).
    \item We conduct simulations for various unseen tasks and compare our proposed MoE-FLAN-SC architecture with three state-of-the-art models, which are the FLAN-T5~\cite{Chung2024}, LM-adapted T5 model~\cite{Raffel2020}, and a conventional approach that uses 8-bit unicode transformation format (UTF-8) encoding and Turbo coding. Simulation results show that our proposed MoE-FLAN-SC architecture achieves a higher average accuracy than the baseline models. It achieves a 3.83\% increase in average accuracy under high SNR conditions without increasing the FLOPs per token. We also conduct ablation study under different numbers of experts and extractor modules in FEM. 
\end{itemize}

This paper is organized as follows. Section~\ref{Sec:framework} introduces the system model and presents the tasks and prompts considered in our model. Section~\ref{Sec:architecture} presents the architecture of the proposed model for each component, including the encoder, FEM, and decoder. We also present the two-phase training algorithm. Results for performance evaluation and comparison are presented in Section~\ref{Sec:result}. Conclusion is given in Section~\ref{Sec:conclusion}.

\textit{Notations}: We use upper-case boldface letters to denote matrices and lower-case boldface letters to denote column vectors. $\mathbb{R}$ and $\mathbb{C}$ denote the sets of real and complex numbers, respectively. $ \mathbf{A}^T$ denotes the transpose of matrix $\mathbf{A}$. $ \mathbf{A}[\mathcal{I}, :] $ returns the submatrix of matrix $ \mathbf{A} $ with the rows chosen from an index set $ \mathcal{I} $. $\mathbb{E}[\cdot]$ represents statistical expectation, and $\mathbf{I}$ denotes the identity matrix.

\section{System Model and Task Description}
\label{Sec:framework}

As illustrated in Fig.~\ref{Fig:system_model}, the proposed MoE-FLAN-SC architecture is trained on a mixture of text-specific datasets using natural language prompts. The model consists of a transmitter and a receiver. The transmitter includes an MoE-based semantic encoder, a multi-task FEM, and a filter module. The MoE-based semantic encoder generates feature representations from the input data. The multi-task FEM produces a mask to filter out the task-irrelevant features. The filter module then removes those masked feature vectors. At the receiver, an MoE-based semantic decoder is used to process the received message for the intended task.

The MoE-based semantic encoder in the transmitter maps a variable-length sentence, denoted as $\mathbf{x}$, into a feature matrix, represented as $\mathbf{Z} \in \mathbb{R}^{N\times D}$, where $N$ is the number of feature vectors and $D$ is the dimension of a feature vector. We have
\begin{align}
    \mathbf{Z} = \mathcal{F_S}(\mathbf{x};\mathbf{\theta_{SE}}), 
\end{align}
where $\mathcal{F_S}(\cdot;\mathbf{\theta_{SE}})$ represents the MoE-based semantic encoder with parameters set $\mathbf{\theta_{SE}}$.
To mask the irrelevant features and reduce the transmission overhead, a multi-task FEM is used to compress the number of features. Given the feature matrix $\mathbf{Z}$ and SNR, the multi-task FEM generates a mask $\mathbf{m} \in \{0,1\}^{N}$, which is given by
\begin{align}
    \mathbf{m} = \mathcal{F_E}(\mathbf{Z}, \gamma;\mathbf{\theta_{E}}), 
\end{align}
where $\mathcal{F_E}(\cdot;\mathbf{\theta_{E}})$ represents the multi-task FEM with parameters set $\mathbf{\theta_{E}}$ and $\gamma$ is the SNR obtained via channel output feedback~\cite{Kurka2020}. Let $m_i$ denote the $i$th element of the mask $\mathbf{m}$. We have $m_i=1$ if the $i$th feature vector is retained and $m_i=0$ otherwise.
We denote the number of retained features as $N_{\text{c}}=\mathbf{1}_{N}^{T} \mathbf{m}\leq N$, where $\mathbf{1}_{N}$ is an all-ones column vector with $N$ elements. Here, we define the compression ratio as $\rho \triangleq N_{\text{c}}/N$. Each feature vector in the masked feature matrix $\mathbf{\tilde{Z}}\in \mathbb{R}^{N \times D}$ is defined as the multiplication of the corresponding element in the mask and the feature vector, i.e., $\mathbf{\tilde{Z}}[i,:]=m_i\mathbf{Z}[i,:]$. The masked feature matrix $\mathbf{\tilde{Z}}$ is sent to the filter module to remove the masked features. Define $\mathcal{R}=\{i~|~m_i=1,~i=1,…,N\}$ as the set of indices for the retained features. The compressed feature matrix $\hat{\mathbf{Z}}$ is given by
\begin{align}
    \mathbf{\hat{Z}} = \mathbf{\tilde{Z}}[\mathcal{R},:]\in \mathbb{R}^{N_{\text{c}}\times D}.
\end{align}
Thus, $\mathbf{\hat{Z}}$ is a submatrix formed by selecting the retained feature vectors from matrix $\mathbf{\tilde{Z}}$.

The compressed feature matrix $\hat{\mathbf{Z}}$ is then reshaped into a stream of encoded symbols $\mathbf{\hat{z}}$, and are transmitted through a wireless channel. The received signal $\mathbf{\hat{y}}$ is given by
\begin{align}
   \mathbf{\hat{y}} = h \mathbf{\hat{z}}+\mathbf{n}, 
\end{align}
where $h\sim \mathcal{CN}(0,1)$ represents the Rayleigh fading channel gain and $\mathbf{n} \sim \mathcal{CN}(\mathbf{0},\sigma^2\mathbf{I})$ denotes the additive white Gaussian noise (AWGN) with zero mean and covariance matrix $\sigma^2\mathbf{I}$. Upon receiving $\mathbf{\hat{y}}$, the stream is reshaped back to a matrix $\hat{\mathbf{Y}}\in\mathbb{R}^{N_{\text{c}} \times D}$. The MoE-based semantic decoder is subsequently used to decode $\hat{\mathbf{Y}}$ into sentence $\mathbf{y}$, representing the response for the intended task. We have
\begin{align}
    \mathbf{y} = \mathcal{G_S}(\mathbf{\hat{Y}};\mathbf{\theta_{SD}}),
\end{align}
where $\mathcal{G_S}(\cdot;\mathbf{\theta_{SD}})$ is the MoE-based semantic decoder with parameters set $\mathbf{\theta_{SD}}$.

To demonstrate the model's zero-shot ability to unseen tasks, we utilize the datasets from P3 and task taxonomy outlined in~\cite{Sanh2022} to train our model. This taxonomy comprises 12 tasks, including multiple-choice question answering (QA), sentiment analysis, and summarization, spanning a total of 62 datasets. We divide the tasks into two groups. The first group, denoted as $\mathcal{T}_{\text{t}}$, contains 8 tasks and is designated for training the multi-task model. The second group, denoted as $\mathcal{T}_{\text{h}}$, contains the remaining 4 tasks and forms a held-out set. It is used to evaluate the model's generalization ability to unseen tasks. All data inputs in our model are formatted as natural language prompts to facilitate generalization. We adhere to the prompt template outlined in P3 to train our model.

\section{Proposed MoE-FLAN-SC Architecture}
\label{Sec:architecture}

\subsection{Architecture of MoE-based Semantic Encoder}
\label{Sec:MoE_encoder}

\begin{figure}[t]
    \centering
    \includegraphics[width=0.44\textwidth]{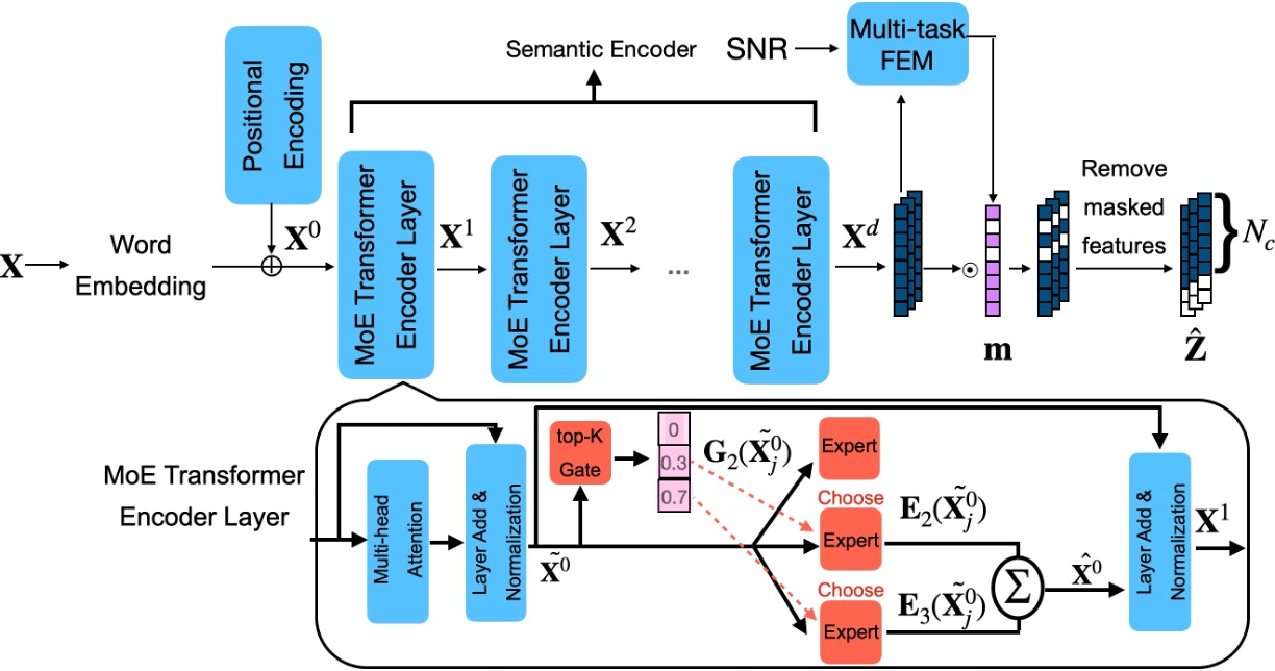}
    \caption{\small The architecture of the MoE-based semantic encoder, where the lower part is the architecture of first MoE Transformer encoder layer. In this figure, the top-2 gate selects $K=2$ experts out of the total $M=3$ experts.}
    \label{Fig:encoder}
\end{figure}

As illustrated in Fig.~\ref{Fig:encoder}, the input sentence $\mathbf{x}$ is initially converted into word embeddings through tokenization. Subsequently, positional embeddings are incorporated into the word embeddings to provide position information for each token. Let $\mathbf{X}^{i}\in \mathbb{R}^{N \times D}$ denote the feature generated by the $i$th encoder layer. The token embeddings are then fed through $E$ layers of the MoE Transformer encoder layer. The output of the final MoE Transformer encoder layer, $\mathbf{X}^{E}$, is also the output of the MoE-based semantic encoder, denoted as $\mathbf{Z}$ in Section~\ref{Sec:framework}.

The architecture of MoE Transformer encoder layer is depicted in the lower part of Fig.~\ref{Fig:encoder}. In the MoE Transformer encoder layer, the features generated from the previous layer $\mathbf{X}^{i-1}$ are first passed through the multi-head self-attention module~\cite{vas2017}, which computes the attention scores within the $i$th MoE Transformer encoder layer, transforming the features into $\tilde{\mathbf{X}}^{i-1}\in \mathbb{R}^{N \times D}$. To enable the model to handle complex characteristics of multiple tasks, we incorporate multiple feed-forward layers~\cite{vas2017}, referred to as experts, in our model. Let $\mathbf{E}_i(\cdot): \mathbb{R}^{D} \rightarrow \mathbb{R}^{D}$ represent the transformation function of the $i$th expert, for $i = 1, \ldots, M$. Instead of utilizing all the results from the experts, we sparsely activate $K$ experts from the total of $M$ experts, where $K < M$. The selection of experts and their corresponding weights are determined by a top-$K$ gate, which selects experts based on the feature matrix $\tilde{\mathbf{X}}^{i-1}$. For each row vector of $\tilde{\mathbf{X}}^{i-1}$, denoted as  $\tilde{\mathbf{X}}^{i-1}[j,:] \in \mathbb{R}^{1 \times D}$, $j=1 ,\ldots, N$, the gating function is given by:
\begin{align}
    \mathbf{G}_K(\tilde{\mathbf{X}}^{i-1}[j,:])=\text{Softmax}(\text{TopK}(\tilde{\mathbf{X}}^{i-1}[j,:]W_g)) \in \mathbb{R}^{1\times M},
\end{align}
where $W_g \in \mathbb{R}^{D \times M}$ is the linear weight matrix of the gate, and TopK$(\cdot)$ retains the largest $K$ values, setting the other elements to negative infinity. Let $\mathbf{E}(\tilde{\mathbf{X}}^{i-1}[j,:])=$  $[\mathbf{E}_1(\tilde{\mathbf{X}}^{i-1}[j,:]),$ $\ldots, \mathbf{E}_M(\tilde{\mathbf{X}}^{i-1}[j,:])]  \in \mathbb{R}^{M \times D}$ denote the concatenated results of the expert transformations. For $j=1,\ldots,N$, the weighted sum for the selected experts is computed as:
\begin{align}
    \hat{\mathbf{X}}^{i-1}[j,:] = \mathbf{G}_K(\tilde{\mathbf{X}}^{i-1}[j,:]) \mathbf{E}(\tilde{\mathbf{X}}^{i-1}[j,:]) \in \mathbb{R}^{1 \times D}.
\end{align}
The result of $\hat{\mathbf{X}}^{i-1}$ is then linearly added with $\tilde{\mathbf{X}}^{i-1}$ and normalized to produce the output of the $i$th MoE Transformer encoder layer $\mathbf{X}^{i}$.

\subsection{Architecture of Multi-Task Feature Extractor Module}
\label{Sec:FEM}
\begin{figure}[t]
\centering
\includegraphics[width=0.48\textwidth]{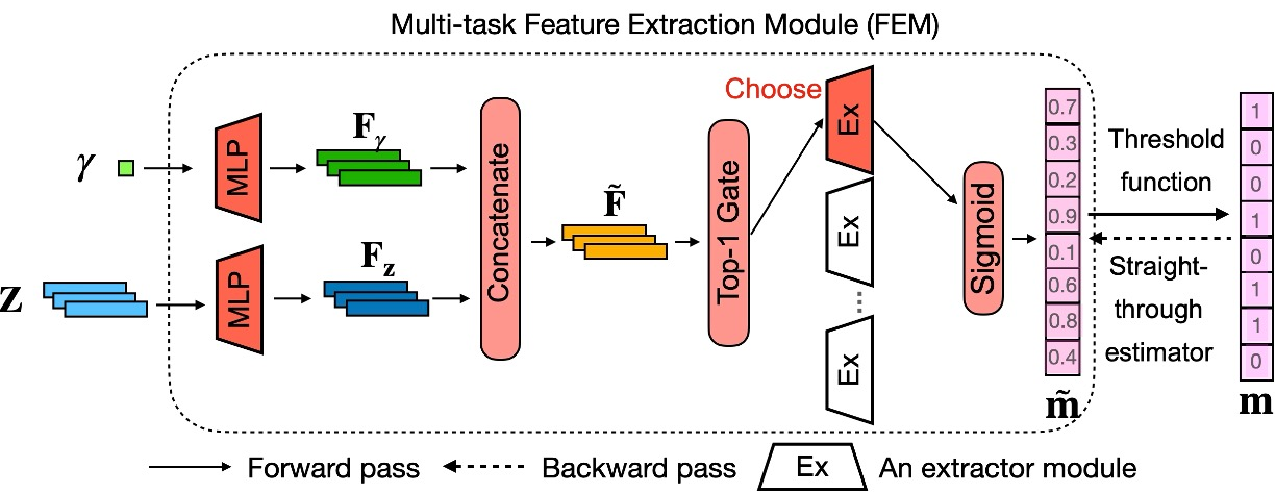}
\caption{\small The architecture of the multi-task feature extraction module.}
\label{Fig:FEM}
\end{figure}

In task-oriented communication systems, not all features are relevant to the intended task. Therefore, a multi-task FEM is used to filter out irrelevant information, retaining only the essential features necessary for the task. As illustrated in Fig.~\ref{Fig:FEM}, the multi-task FEM takes into account the SNR and the features generated by the semantic encoder $\mathbf{Z}$ to generate a binary mask, $\mathbf{m}\in \{0,1\}^{N}$, for each feature. 
Both inputs are processed through a multilayer perceptron (MLP) to generate representations of the SNR and the features, denoted as $\mathbf{F}_{\gamma}$ and $\mathbf{F}_{z}$, respectively. To jointly consider both inputs, we concatenate these feature representations as $\tilde{\mathbf{F}}=\text{cat}(\mathbf{F}_{\gamma}, \mathbf{F}_{z})$. Since each task has distinct characteristics, the multi-task FEM should capture different feature characteristics and differentiate relevant features. To enhance feature identification across a wide range of tasks, we employ a total of $B$ extractor modules within the multi-task FEM. The matrix $\tilde{\mathbf{F}}$ is then passed through one of these $B$ extractor modules followed by a sigmoid activation function. The selection of the extractor module is determined by a Top-1 gate $\mathbf{G}_1(\cdot)$. Let $\mathbf{M}_{k}(\cdot)$ denote the linear transformation of the $k$th extractor module. The mask $\tilde{\mathbf{m}}= [\tilde{m}_1, \ldots, \tilde{m}_N]\in{[0,1]}^{N}$ is generated as
\begin{align}
    \tilde{\mathbf{m}} = \text{Sigmoid}(\mathbf{M}_{k}(\tilde{\mathbf{F}})) ,~ \text{where}~k=\arg\max \mathbf{G}_1(\tilde{\mathbf{F}}).
\end{align}
This mask indicates the importance of each feature, with values ranging between 0 and 1. To determine whether a feature should be retained or discarded, a constant threshold $\bar{t}$ is applied. The final mask $\mathbf{m}= [m_1, \ldots, m_N]$ is determined as follows:
\begin{align}
    m_i = \begin{cases}
        1, & \tilde{m}_i \geq \bar{t},\\
        0, & \text{otherwise}.
    \end{cases}
    \label{Eq:STE}
\end{align}
In the forward pass, the threshold function (\ref{Eq:STE}) is applied. However, in the backward pass, the threshold function (\ref{Eq:STE}) is non-differentiable, which brings a challenge for backpropagation. To address this issue, we use the straight-through estimator~\cite{Bengio2013} during the backward pass to approximate the gradient. 

\subsection{Architecture of MoE-based Decoder}
\label{Sec:MoE_decoder}
\begin{figure}[t]
\centering
\includegraphics[width=0.41\textwidth]{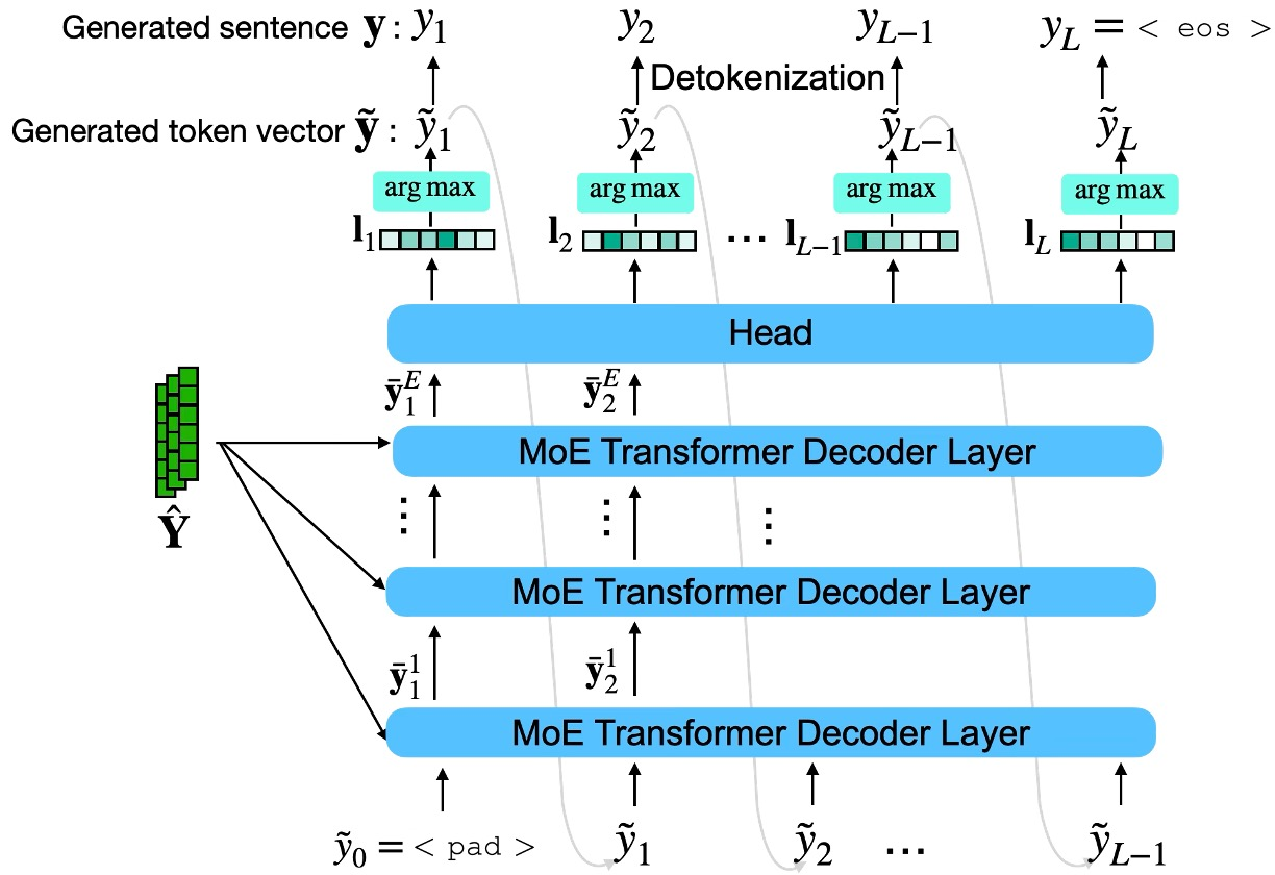}
\caption{\small The architecture of the MoE-based semantic decoder.}
\label{Fig:decoder}
\end{figure}

The architecture of MoE-based decoder is shown in Fig.~\ref{Fig:decoder}. It consists of a stack of MoE Transformer decoder layers, followed by a dense layer and a softmax layer, jointly referred to as the head. Specifically, the MoE Transformer decoder layer maintains the same architecture as the vanilla decoder layer in the Transformer model. The only difference is that the linear layer within the feed-forward network is replaced by MoE and a gating mechanism, as described in Section~\ref{Sec:MoE_encoder}. To generate the $j$th word in output $\mathbf{y}$, denoted as $y_j$, the MoE Transformer decoder layers take the received feature matrix $\hat{\mathbf{Y}} \in \mathbb{R}^{N_{\text{c}}\times D}$ and the previous generated token $\tilde{y}_{j-1}$ as input, where the initial token $\tilde{y}_0$  is designated as the special $<\texttt{pad}>$ token. These inputs are processed to generate $\mathbf{\bar{y}}_{j}^{i}$ from the $i$th MoE Transformer decoder layer for the $j$th word. After passing through a total of $E$ MoE Transformer decoder layers, the head layer maps the final output $\mathbf{\bar{y}}_{j}^{E}$ to a probability vector, denoted as $\mathbf{l}_j \in [0,1]^{V}$, where $V$ represents the number of tokens in the token vocabulary. Each element in vector $\mathbf{l}_j$ represents the probability of taking a specific token in the token vocabulary. The token $\tilde{y}_j$ is determined by selecting the one with the highest likelihood, given as $\tilde{y}_j= \arg\max_{k = 1, \ldots, V} l_{j, k}$, where $l_{j,k}$ is the $k$th element  of vector  $\mathbf{l}_j$. The decoder auto-regressively generates the token $\tilde{y}_j$ until it produces the index of a special $<\texttt{eos}>$ token, thereby constructing the generated token vector $\mathbf{\tilde{y}}=[\tilde{y}_1,\dots,\tilde{y}_L]\in \{1, \ldots, V\}^{L}$ with $L$ tokens. By detokenizing $\mathbf{\tilde{y}}$, we can obtain the generated sentence $\mathbf{y}$ with $L-1$ words since the $<\texttt{eos}>$ token indicates the end of sentence and does not correspond to a word.

\subsection{Proposed MoE-FLAN-SC Training Algorithm}
\label{Sec:algorithm}
MoE-FLAN-SC is designed to achieve two objectives: learning general representations across multiple tasks, and ensuring that the FEM filters out task-irrelevant features. To jointly optimize the model with two different objectives, we propose a two-phase training algorithm to train the proposed MoE-FLAN-SC. The proposed training algorithm is shown in Algorithm~\ref{alg:train}. In the first phase, we train the MoE-based semantic encoder and decoder without applying any compression, aiming to establish a model capable of capturing the essential features. To optimize the model, we use the cross-entropy loss function. Let $\mathbf{q} = [q_1, \ldots, q_L]\in \{1,\ldots,V\}^{L}$ denote the tokens of the target sentence. For the $i$th token in $\mathbf{q}$, denoted as $q_i$, the one-hot vector representation is represented as $\mathbf{o}_i \in \{0,1\}^{V}$, where only the $q_i$th element is equal to 1 and all other elements are set to zero. The cross-entropy loss is as follows
\begin{align}
    \mathcal{L}_{\text{CE}} &= -\frac{1}{L}\sum_{i=1}^{L}\sum_{j=1}^{V}o_{i,j}\log(l_{i,j}),
    \label{Eq:crossentropyloss}
\end{align}
where $o_{i,j}$ is the $j$th element of vector $\mathbf{o}_i$.
In the second phase, we load the trained parameters from the first phase and incorporate the multi-task FEM. To minimize the number of transmit symbols, we introduce the compression ratio as an additional loss function term. It is given by
\begin{align}
    \mathcal{L}_{\rho}=\frac{N_{\text{c}}}{N}.
    \label{Eq:compressionloss}
\end{align}
The model is then trained to jointly minimize the cross-entropy loss $ \mathcal{L}_{\text{CE}}$ and the compression loss $\mathcal{L}_{\rho}$. This joint optimization facilitates the model in achieving high performance in generating the target tokens while filtering out irrelevant features, thereby reducing the number of transmit symbols.
\begin{algorithm}[t]
\small
\caption{Proposed MoE-FLAN-SC Training Algorithm}\label{alg:train}
\begin{algorithmic}[1]
\State \textbf{Input:} Training datasets for each task $t$ in $\mathcal{T}_{\text{t}}$, including data and labels; Number of epochs for the first and second phases, $N_1$ and $N_2$; learning rate $\eta$; regularization weights $\omega_1$ and $\omega_2$.
\State \textbf{First Phase:}
\State Fixed the parameters in FEM.
\For{$i \gets 1$ \textbf{to} $N_1$} 
    \For{$t \in \mathcal{T}_{\text{t}}$}
        \State Load input data $\mathbf{x}$ and labels for task $t$.
        \State Sample the channel gain coefficient $h$ and SNR value.
        \State Perform forward propagation.
        \State Calculate the loss $\mathcal{L}_1(\mathbf{x}) \mathrel{:=} \mathcal{L}_{\text{CE}}$ based on (\ref{Eq:crossentropyloss}).
        \State Update parameters $\{\mathbf{\theta_{SE}}, \mathbf{\theta_{SD}}\}$ \text{using }$\mathcal{L}_1(\mathbf{x})$.
    \EndFor
\EndFor
\State \textbf{Second Phase}: 
\State Load parameter sets trained in the first phase to the model.
\For{$i \gets 1$ \textbf{to} $N_2$} 
    \For{$t \in \mathcal{T}_{\text{t}}$}
        \State Load input data $\mathbf{x}$ and labels for task $t$.
        \State Sample the channel gain coefficient $h$ and SNR value.
        \State Perform forward propagation.
        \State Calculate the loss $\mathcal{L}_2(\mathbf{x})\mathrel{:=} \omega_1 \mathcal{L}_{\text{CE}}+\omega_2\mathcal{L}_{\rho}$ based on (\ref{Eq:crossentropyloss}) and (\ref{Eq:compressionloss}).
        \State Update parameters $\{\mathbf{\theta_{SE}}, \mathbf{\theta_{E}},\mathbf{\theta_{SD}}\}$ \text{using }$\mathcal{L}_2(\mathbf{x})$.
    \EndFor
\EndFor
\State \textbf{Output}: The optimized parameters $\{\mathbf{\theta_{SE}^{*}}, \mathbf{\theta_{E}^{*}}, \mathbf{\theta_{SD}^{*}}\}$.
\end{algorithmic}
\end{algorithm}

\section{Performance Evaluation}
\label{Sec:result}

The MoE-based semantic encoder and decoder of the MoE-FLAN-SC are initialized with the FLAN-T5-Base model\footnote{The public checkpoints for FLAN-T5 models are available at: \url{https://github.com/google-research/t5x/blob/main/docs/models.md#flan-t5-checkpoints}}~\cite{Chung2024} such that the number of Transformer layers $E$ is set to 12, and the dimension for each feature vector $D$ is set to 768. Unless stated otherwise, the number of experts $M$ is set to 10, the number of selected experts $K$ is set to 1, and the number of extractor modules $B$ is set to 10. All the experts are initialized with the same parameters as in the original feed-forward layer in FLAN-T5-Base model. The models are trained on P3~\cite{Sanh2022}, a collection of datasets with prompt templates. For the settings of the training procedure, we use AdamW optimizer~\cite{Losh2019} with learning rate $\eta=10^{-5}$ and batch size equal to 120. The token selection threshold $\bar{t}=0.5$ and the regularization weights $(\omega_1, \omega_2)=(10^{3}, 10)$. In our simulations, we use the exact match to evaluate the accuracy by measuring the proportion of generated sentences that exactly match the targets. We consider the following baseline models for comparison:
\begin{enumerate}
    \item \textbf{FLAN-SC model}~\cite{Chung2024}: This model is initialized with FLAN-T5-Base model. It only includes one expert in the Transformer layer and does not include the  gating network.
    \item \textbf{LM-adapted SC model}\footnote{The public checkpoints for LM-adapted-T5 models are available at: \url{https://github.com/google-research/text-to-text-transfer-transformer/blob/main/released_checkpoints.md\#lm-adapted-t511lm100k}}~\cite{Raffel2020}: This model is initialized with LM-adapted T5 model and is trained without instruction-tuning. It includes only a single expert in each Transformer layer and does not have a gating network. We initialize the model with the ``XL" size of the LM-adapted T5 model, which consists of $E=24$ Transformer layers. The dimension $D$ is equal to $2048$.
    \item \textbf{UTF-8+Turbo coding}: The UTF-8 encoding and Turbo coding are employed for source coding and channel coding, respectively. The code rate for channel coding is set to $\frac{1}{2}$. The FLAN-T5-Base model is utilized at the receiver to perform the intended tasks based on the decoded message.
\end{enumerate}
 \begin{figure}[t]
    \centering
    
    \begin{subfigure}[b]{0.24\textwidth}
        \centering
        \includegraphics[width=\textwidth]{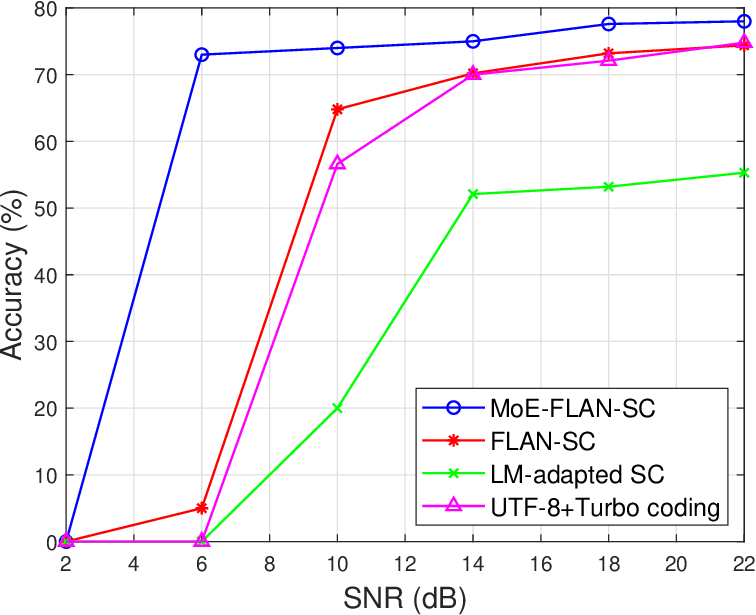}
        \caption{Sentence Completion}
        \label{fig:suba}
    \end{subfigure}
    \hfill
    \begin{subfigure}[b]{0.24\textwidth}
        \centering
        \includegraphics[width=\textwidth]{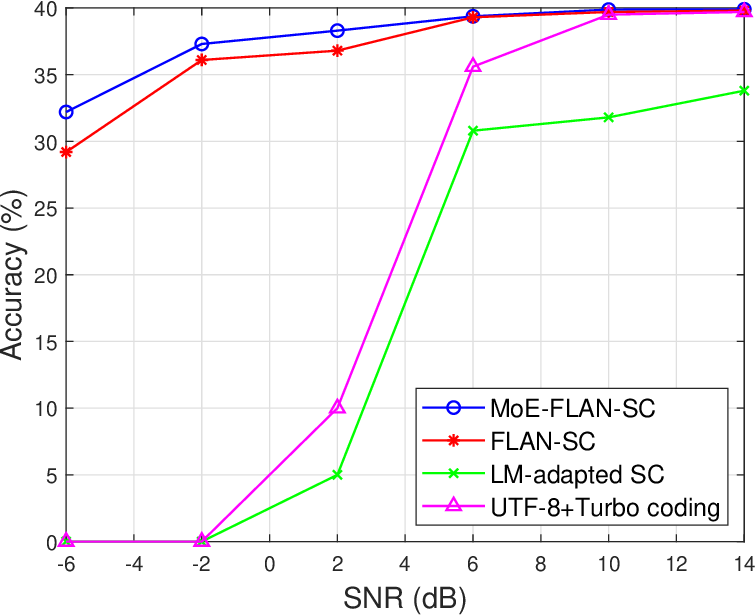}
        \caption{Natural Language Inference}
        \label{fig:subb}
    \end{subfigure}
    \hfill
    \begin{subfigure}[b]{0.24\textwidth}
        \centering
        \includegraphics[width=\textwidth]{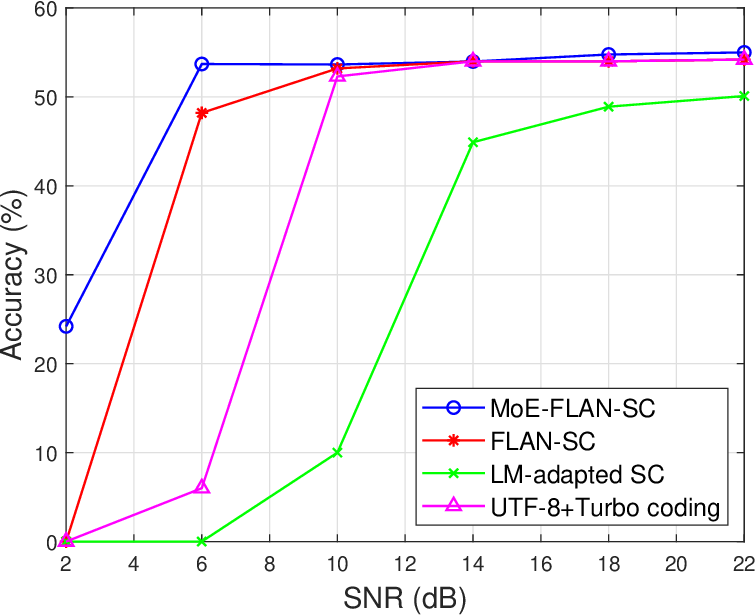}
        \caption{Coreference Resolution}
        \label{fig:subc}
    \end{subfigure}
    \hfill
    \begin{subfigure}[b]{0.24\textwidth}
        \centering
        \includegraphics[width=\textwidth]{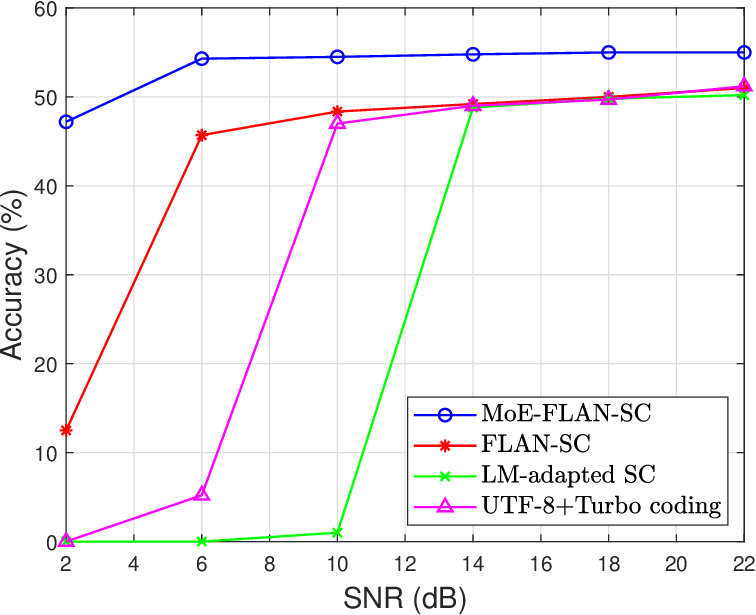}
        \caption{Word Sense Disambiguation}
        \label{fig:subd}
    \end{subfigure}
    
    \caption{The accuracy of the held-out tasks in $\mathcal{T}_{\text{h}}$ versus SNR. (a) Sentence completion task, (b) natural language inference task, (c) coreference resolution task, (d) word sense disambiguation task.}
    \label{fig:held-out}
\end{figure}

\begin{figure}[t]
        \centering
        \includegraphics[width=0.48\textwidth]{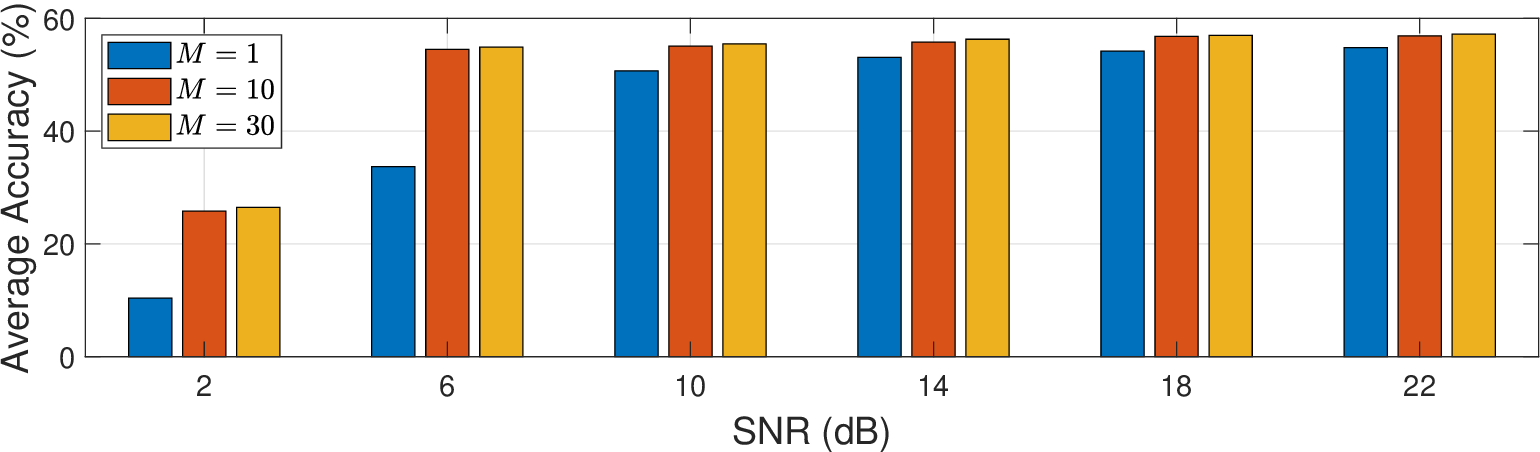}
        \caption{The average accuracy of all four held-out tasks versus SNR under different number of experts $M$.}
        \label{fig:num_experts}
    \end{figure}

In Fig.~\ref{fig:held-out}, we show the performance of different baseline models versus SNR for four held-out tasks. Across all tasks, our proposed MoE-FLAN-SC consistently outperforms FLAN-T5, while FLAN-T5 outperforms the other baseline models. This demonstrates that instruction-tuning enhances the performance of the model across different tasks, and integrating the MoE architecture into FLAN-T5 leads to significant performance gains. Among those four unseen tasks, MoE-FLAN-SC achieves an average accuracy of 56.9\%, compared to 54.8\% for FLAN-SC when SNR = 22 dB, representing a 3.83\% accuracy improvement in high SNR region. The results highlight the superiority of MoE-FLAN-SC over FLAN-T5. In Fig.~\ref{fig:num_experts}, we show that when the number of experts $M$ increases in our proposed MoE-FLAN-SC architecture, the performance will eventually saturate.

\begin{figure}[t]
    \centering
    \begin{subfigure}[b]{0.24\textwidth}
        \centering
        \includegraphics[width=\textwidth]{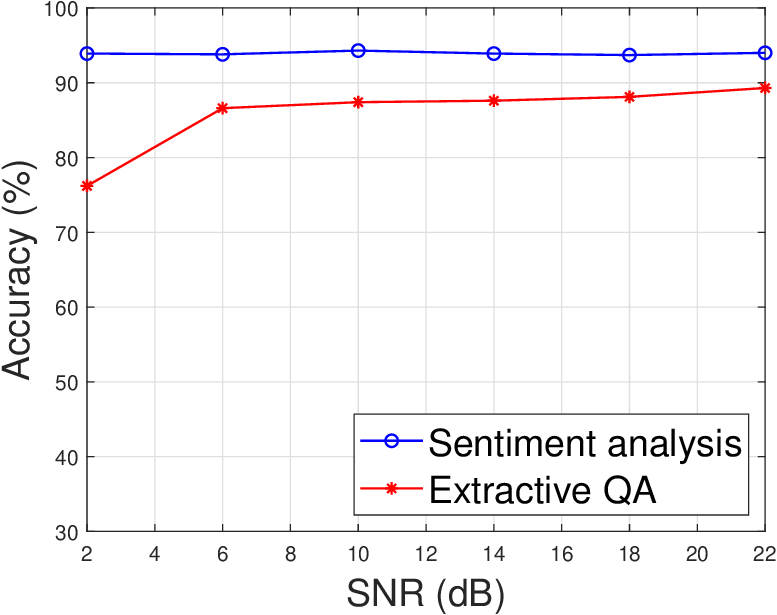}
        \caption{}
    \end{subfigure}
    \hfill 
    \begin{subfigure}[b]{0.24\textwidth}
        \centering
        \includegraphics[width=\textwidth]{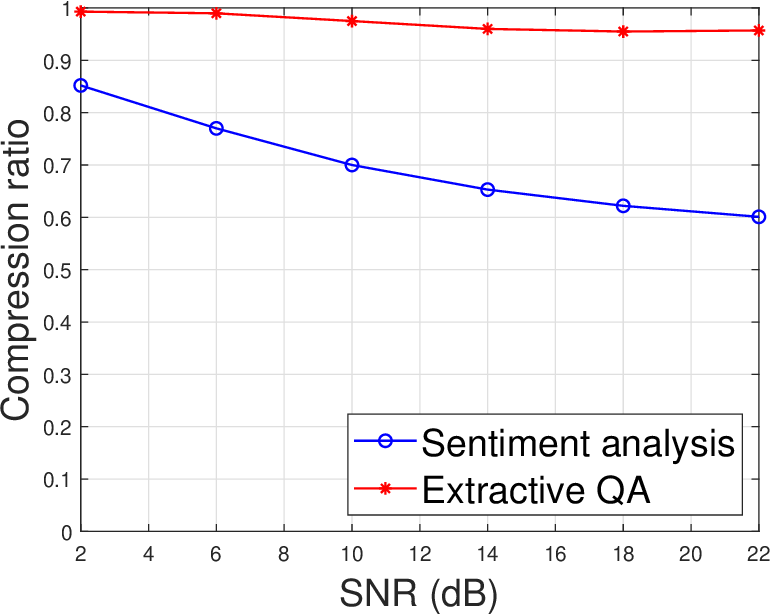}
        \caption{}
    \end{subfigure}
    \caption{ (a) The accuracy and (b) compression ratio versus SNR for sentiment analysis and extractive QA tasks.}
    \label{Fig:compress}
\end{figure}

\begin{figure}[t]
    \centering
    \begin{subfigure}[b]{0.24\textwidth}
        \centering
        \includegraphics[width=\textwidth]{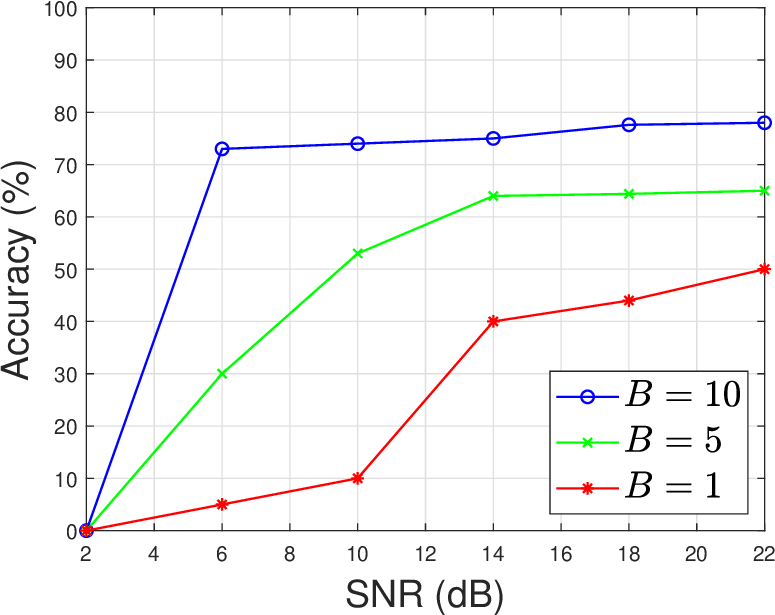}
        \caption{}
    \end{subfigure}
    \hfill 
    \begin{subfigure}[b]{0.24\textwidth}
        \centering
        \includegraphics[width=\textwidth]{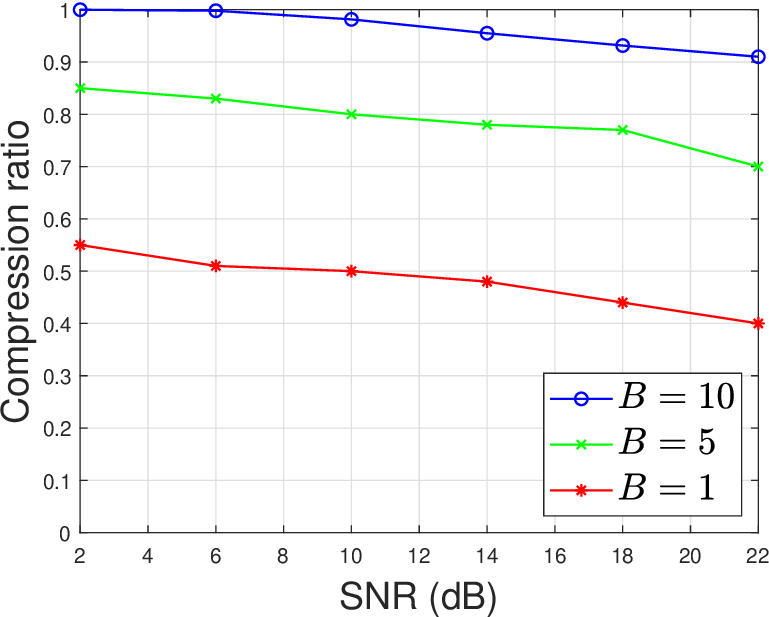}
        \caption{}
    \end{subfigure}
    \caption{(a) The accuracy and (b) compression ratio versus SNR for sentence completion task under number of extractor modules $B$.}
    \label{Fig:num_MLP}
\end{figure}

To evaluate the effectiveness of multi-task FEM in our proposed architecture, Fig.~\ref{Fig:compress} presents the accuracy and compression ratio for sentiment analysis and extractive QA tasks. When the SNR decreases, the compression ratio increases, indicating that lower SNR requires more redundancy to maintain the performance. At higher SNR, less redundancy is needed, allowing higher compression without loss of accuracy. The results also show different compression ratios required for different tasks. QA requires the entire paragraph to find the answer. It requires high compression ratios even at high SNR region. On the other hand, sentiment analysis is a binary task. It allows higher compression when the SNR increases. These results show that the multi-task FEM can effectively adapt the transmit symbols to different tasks and SNR conditions. In Fig.~\ref{Fig:num_MLP}, we present the results of using different number of extractor modules $B$ in the FEM. Results show that utilizing multiple extractor modules in the FEM can improve the accuracy of unseen tasks.

Finally, in Table~\ref{tab:computation}, we present the results of the total number of parameters and FLOPs per token for different models. The use of MoE increases the total number of parameters count on the order of $O(ME)$, where $M$ is the number of experts per layer and $E$ is the number of layers, while the FLOPs per token remain the same since we only select one expert to process the input feature vector. This demonstrates that the MoE architecture enables scaling the model without increasing the computational cost.
\begin{table}[t]
    \caption{\centering\textsc{Total Number of Parameters and FLOPs per Token of Different Models}}
    \centering
    \resizebox{0.8\columnwidth}{!}{
    \begin{tabular}{|c| c| c|}
        \hline
        \textbf{Model} & \textbf{Total Parameters} & \textbf{FLOPs per token}\\
        \hline
        MoE-FLAN-SC & 1.27B & 300M \\
        \hline
        FLAN-SC & 250M &  300M\\
        \hline
        LM-adapted SC & 3B &  3.6G\\
        \hline
    \end{tabular}}
    \label{tab:computation}
\end{table}

\section{Conclusion}
\label{Sec:conclusion}
In this paper, we proposed an MoE-FLAN-SC architecture for multi-task SC systems. The proposed architecture integrated an MoE into an instruction-tuned LLM and is capable to perform zero-shot learning on unseen tasks. Simulation results showed that our proposed architecture provides good performance across four unseen tasks. Results also demonstrated that our proposed MoE-FLAN-SC architecture outperforms FLAN-SC, LM-adapted SC, and UTF-8+Turbo coding, achieving an average accuracy improvement of 3.83\% on unseen tasks compared to FLAN-SC. Furthermore, the proposed multi-task FEM can effectively optimize the number of transmit symbols based on task requirements and channel conditions. For future work, we plan to extend the proposed architecture to improve the zero-shot performance in multimodal communication scenario.

\bibliographystyle{ieeetr}
\bibliography{refs}
\end{document}